\documentclass[lettersize,journal]{IEEEtran}
\usepackage{amsmath,amsfonts}
\usepackage{algorithmic}
\usepackage{algorithm}
\usepackage{array}
\usepackage[caption=false,font=normalsize,labelfont=sf,textfont=sf]{subfig}
\usepackage{textcomp}
\usepackage{stfloats}
\usepackage{url}
\usepackage{verbatim}
\usepackage{graphicx}
\usepackage[square, comma, sort&compress, numbers]{natbib}
\usepackage{cases}
\usepackage{amssymb}
\usepackage{amsthm}
\usepackage{hyperref}

\usepackage{array}
\usepackage{textcomp}
\usepackage{epsfig}
\usepackage{caption}
\usepackage{epstopdf}
\usepackage{float}
\usepackage{color}
\usepackage{threeparttable}
\usepackage{booktabs}
\usepackage[inkscapelatex=false]{svg}
\theoremstyle{remark}

\begin{document}

\title{Coordinated Spectral Efficiency Prediction \\
	 for Real-World 5G CoMP Systems}

\author{Zhixing Chen, Zhaoyu Fan, Yang Li, Yibin Kang, Qi Yan, Qingjiang Shi

}

\author{Zhixing Chen, Zhaoyu Fan, Yang Li, Yibin Kang, Qi Yan, and Qingjiang Shi 
	\thanks{Zhixing~Chen and Zhaoyu Fan are with the School of Software Engineering, Tongji University, Shanghai, China (e-mail: zhixing@tongji.edu.cn, zhaoyu\_fan@tongji.edu.cn). These two authors contributed equally to this work.}
	
	\thanks{Yang Li is with the Shenzhen Research Institute of Big Data, Shenzhen, China (e-mail: liyang@sribd.cn).}
	
	\thanks{Yibin Kang and Qi Yan are with the Networking and User Experience Lab, Huawei Technologies, Shenzhen, China (e-mail: kangyibin@huawei.com, yanqi1@huawei.com).}
	
	\thanks{Qingjiang Shi is with the School of Software Engineering, Tongji University,
	Shanghai, China, and also with Shenzhen Research Institute of Big Data, Shenzhen, China (e-mail: shiqj@tongji.edu.cn).}
	}


\maketitle

\begin{abstract}
Coordinated multipoint (CoMP) systems incur substantial resource consumption due to the management of backhaul links and the coordination among various base stations (BSs). 
Accurate prediction of coordinated spectral efficiency (CSE) can guide the optimization of network parameters, resulting in enhanced resource utilization efficiency.
However, characterizing the CSE is intractable due to the inherent complexity of the CoMP channel model and the diversity of the 5G dynamic network environment, which poses a great challenge for CSE prediction in real-world 5G CoMP systems.
To address this challenge, in this letter, we propose a data-driven model-assisted approach.
Initially, we leverage domain knowledge to preprocess the collected raw data, thereby creating a well-informed dataset. 
Within this dataset, we explicitly define the target variable and the input feature space relevant to channel statistics for CSE prediction.
Subsequently, a residual-based network model is built to capture the high-dimensional non-linear mapping function from the channel statistics to the CSE. 
The effectiveness of the proposed approach is validated by experimental results on real-world data.
\end{abstract}


\section{Introduction}
\IEEEPARstart{C}{oordinated} multipoint (CoMP) is one of the promising concepts for 5G communications. 
By interconnecting multiple base stations (BSs) through high-capacity backhaul links, CoMP can serve users in a coordinated approach.
This coordination mitigates intercell interference, expands coverage, and enhances data rates for users at cell edges 
\cite{irmer2011coordinated,yang2013we,karakayali2006network}. 
However, the CoMP systems incur substantial resource consumption due to the management of backhaul links and the coordination among various BSs.
Consequently, it is imperative to optimize network parameters, such as the modulation and coding scheme (MCS) and the selection of coordinating BSs, to improve resource utilization efficiency \cite{li2014energy}. 
Coordinated spectral efficiency (CSE) represents the aggregate spectral efficiency (SE) achieved through the coordinated efforts of multiple BSs.
It serves as a key performance indicator for communication quality, thereby providing a foundation for optimizing the network parameters of CoMP systems.


In real-world downlink CoMP systems, the CSE is obtained by calculating the ratio of the received downlink traffic at user equipment (UE) to the occupied bandwidth given the setting of the network parameters. 
However, if the CSE corresponding to any setting of the network parameters can be accurately predicted, these parameters can be optimized purposefully based on the predicted CSE, thus achieving the goal of enhancing resource utilization efficiency \cite{li2022real}. 
Nevertheless, it is intractable to characterize the CSE due to the inherent complexity of the CoMP channel model and the diversity of the 5G dynamic network environment.
This poses a great challenge for the CSE prediction in real-world 5G CoMP systems.

Recently, the generalization and robust fitting capabilities of machine learning methods have sparked growing interest in their application for performance prediction in wireless communications so as to enhance various dimensions of network functionality \cite{gijon2022data,xing2022spectrum}.
The authors of \cite{gijon2022data} evaluated the accuracy of throughput predictions, exploring a variety of machine learning techniques and artificial neural network-based ensemble approaches.
In \cite{xing2022spectrum}, the authors proposed a neural network that leverages channel statistics to predict SE in downlink MIMO transmissions, which highlights the promising capabilities of machine learning for performance prediction.
However, due to the distinct operation and implementation complexity of the CoMP system, these methods cannot be applied to predict the CSE in CoMP systems. The CSE of a CoMP system remains predominantly at the theoretical analysis stage \cite{yang2013we,ali2018downlink,kim2023deep}.
For instance, the authors of \cite{ali2018downlink} investigated the application of CoMP to non-orthogonal multiple access-based multi-cell downlink transmissions to improve the CSE of the multi-cell networks.
In \cite{kim2023deep}, the authors proposed a novel technique employing deep neural networks for dynamic BS selection and power allocation in the CoMP transmission, which offers a viable strategy for the CSE enhancement.
\textit{Despite the above works, to the best of our knowledge, no existing research has focused on the CSE prediction based on real-world CoMP system data.
}

 The main contributions of this letter are two-fold.
 \begin{itemize}
 	\item \textit{Domain Knowledge-Based Efficient Data Cleaning and Feature Space Definition:}
 	
	 The CSE should be higher than the non-coordinated SE, in line with the theoretical principles of CoMP operation. To ensure a reliable data foundation, we exclude data entries that significantly deviate from this principle from the collected real-world samples. 
	 Moreover, we explicitly define the input feature space relevant to channel statistics, and the target variable, facilitating the training of the predict model and enhancing the interpretability.

 	\item \textit{High-Generalization CSE Prediction Model: }
 	
 	We develop a residual-based CSE prediction neural network (CSEPNN) to fit the non-linear mapping function from the channel statistics to the CSE in CoMP systems. We employ supervised learning and calculate the loss using the Huber loss function to train the model. The model's performance is validated through a variety of test cases, with experimental results demonstrating its effective predictive capabilities in real-world 5G CoMP systems.
 \end{itemize}


The remainder of the letter is organized as follows. In Section \ref{sec:background}, we provide an overview of real-world 5G CoMP systems. In Section \ref{sec:data}, we first describe the details of raw data collection, and then perform preprocessing to obtain the dataset. Section \ref{sec:model} presents the structure of the proposed CSEPNN model for CSE prediction. Experimental results and conclusion are provided in Section \ref{sec:simu} and Section \ref{sec:conclusion}, respectively.


\vspace{-15pt}
\section{System Overview} \label{sec:background}

In a real-world 5G downlink CoMP system with coordination of at most two BSs, a UE can be served through two operational modes: Single-Transmission (ST) mode and Joint-Transmission (JT) mode.
In ST mode, only a primary BS transmits the signal to the UE.
In contrast, JT mode involves a secondary BS coordinating with the primary BS to jointly provide services to the UE by transmitting the same signal simultaneously.
In particular, the selection of the primary BS is determined by the BS handover criterion in the 5G network. 
Subsequently, a specific neighboring BS is selected as the secondary according to some predefined rules.

Consider the downlink transmission in JT mode of the CoMP system, where both of the primary and the secondary BSs are equipped with $N_{t}$ transmit antennas and each UE has $N_{r}$ receive antennas. Denote $\mathbf{x}\in\mathbb{C}^{d\times 1}$ as the signal transmitted by both BSs to a certain UE, where $d$ is the number of data streams. The received signal $\mathbf{y}\in\mathbb{C}^{N_{r}\times 1}$ at this UE end can be expressed as
\begin{equation*}
	\mathbf{y} =  \left(  \mathbf{H}_{p}\mathbf{V}_{p}+\mathbf{H}_{s}\mathbf{V}_{s}\right)
	\mathbf{x} + \mathbf{n},  
\end{equation*}
where  $\mathbf{H}_{p}\in\mathbb{C}^{N_{r}\times N_{t}}$ represents the channel from the primary BS to the UE and $\mathbf{V}_{p}\in\mathbb{C}^{N_{t}\times d}$ denotes the downlink beamformer of the primary BS, while $\mathbf{H}_{s}\in\mathbb{C}^{N_{r}\times N_{t}}$ and $\mathbf{V}_{s}\in\mathbb{C}^{N_{t}\times d}$ are the channel and the beamformer of the secondary BS, respectively, $\mathbf{n}\in\mathbb{C}^{N_{r}\times 1}$ represents the unwanted signal, which consists of interference from other BSs plus noise.

Under specific precoding schemes, such as zero-forcing, the achievable rate of the UE can be calculated using the Shannon–Hartley theorem \cite{karakayali2006network}. However, this method yields the theoretical upper limit of throughput for communication networks, as it relies on idealized assumptions. This rate should be interpreted as the system capacity rather than the CSE. 
A data-driven approach for predicting the CSE can offer more accurate throughput estimates with lower time complexity.
Consequently, it warrants further exploration in practical CoMP systems.

\vspace{-2pt}
\section{Data Collection and Preprocessing} \label{sec:data}

\subsection{Data Collection}

A 5G mobile phone serves as a UE to collect data samples in frames during driven testing, while the operation of BSs continually alternates between ST mode and JT mode based on the coordinated massive MIMO technique developed by HUAWEI.
The primary goal of data collection is to accurately capture the current instantaneous channel state information (CSI).
Specifically, in ST mode, the received data volume after primary BS transmits downlink traffic to the UE is measured.
In JT mode, the measurement encompasses not only the received traffic as both the primary and secondary BSs serve the UE, but also the $\mathbf{8}$ highest RSRPs of CSI beams in various directions from each BS, resulting in a total of $\mathbf{16}$ RSRPs.

Moreover, a substantial quantity of metadata is collected to enhance the analytical potential of the raw data samples,  which aids in the development of datasets. This metadata consists of timestamps, locations, and BS identification codes corresponding to the collected data samples.

\vspace{-6.9pt}
\subsection{Data Preprocessing}

In order to enhance the quality and applicability of the data and establish a reliable and effective data foundation for constructing network models, we conduct the following preprocessing steps on the raw data samples:

\begin{itemize}	
	\item[1.]\textit{Second-Level Mean Replacement:}
		Replace the raw data with the mean of all frame data within the same second.
	\item[2.]\textit{SE and CSE Calculation:}
		Calculate the SE and CSE through dividing the mean of received data volume in ST and JT mode within seconds by the occupied bandwidth, respectively.
	\item[3.]\textit{Data Alignment:}
		Align data entries across two modes based on time and BS identification codes.
	\item[4.]\textit{Data Cleaning:}
		Exclude outliers where the CSE in JT mode is significantly lower than the SE in ST mode. Based on domain knowledge of CoMP, the CSE should not be lower than the SE, as both BSs transmit the identical signal to the UE. Such outliers might be caused by abrupt interference during data collection. Accounting for an acceptable margin of error, we remove the data items where the CSE is lower than 85\% of the SE.
\end{itemize}
    
As a result, we complete the data collection and preprocessing, which are depicted on the left side of the framework in Fig. \ref{fig:RSEPNN}.  To train a model that exhibits generalizable and robust properties, we create datasets from various times and locations. The $16$ RSRPs from both the primary BS and the secondary BS in JT mode, together with the SE in ST mode, constitute the input feature space whose dimensions are $\mathbf{17}$, while the CSE serves as the label.


\section{Proposed CSEPNN for Coordinated SE Prediction} \label{sec:model}
\begin{figure*}[htbp]
	\centering
	\includegraphics[width = .95\textwidth]{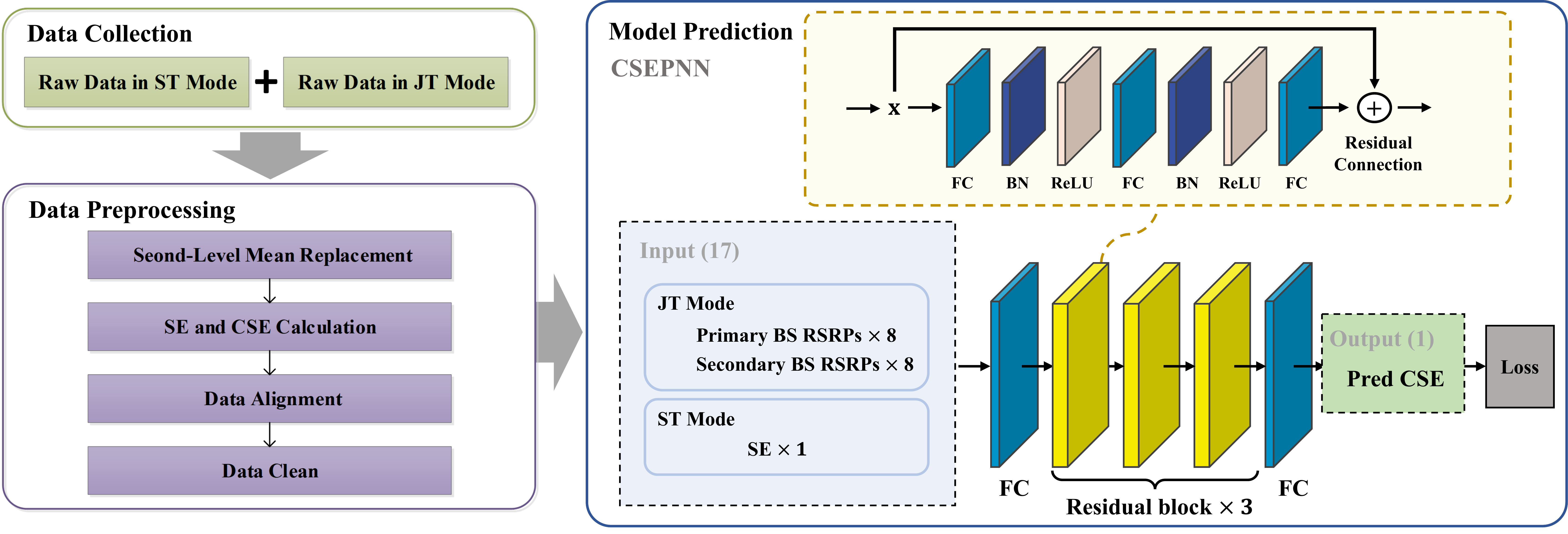}
	\caption{Framework of the CSE prediction approach: data collection, data preprocessing, and model prediction.}
	\label{fig:RSEPNN}
\end{figure*}

In this section, we propose the CSEPNN to predict the CSE. The architecture of the CSEPNN is presented on the right hand in Fig. \ref{fig:RSEPNN}.

The CSEPNN is a deep neural network characterized by the residual structure. As shown in Fig.\ref{fig:RSEPNN}, the CSEPNN takes $17$ real-world collected values as its input, and predicts the CSE as its output. It consists of an individual fully connected (FC) input layer, followed by three residual blocks, and concludes with another individual FC output layer. 
Each FC layer contains $200$ neurons. In a residual block, the number of the three FC layers are $200$, $100$ and $200$, respectively. Between adjacent FC layers, batch normalization (BN) and ReLU activation function are applied sequentially. Furthermore, a residual connection at the end facilitates information skipping across layers, thereby contributing to gradient stability and mitigating the vanishing gradient issue effectively. Therefore, the proposed CSEPNN is capable of efficiently capturing the complex high-dimensional non-linear mapping function from the channel statistics to the CSE.

Applying supervised learning, the Huber loss function \cite{hastie2017elements} is utilized to compute the loss between label and the output. The expression for the Huber loss function is given by:
\begin{equation} \nonumber
    L_{\delta}(y, f(x)) = \begin{cases}
    \frac{1}{2}(y - f(x))^2 & \text{if } |y - f(x)| \leq \delta \\
    \delta (|y - f(x)| - \frac{1}{2}\delta) & \text{otherwise}
    \end{cases}
\end{equation}
where $y$ and $f(x)$ denote the values of label and output, respectively, $\delta$ is a hyperparameter. The function provides a smooth transition between the quadratic behavior of the  mean squared error for small errors and the linear behavior of the mean absolute error (MAE) for large errors, making it advantageous in predicting the CSE.

\section{Experimental Results} \label{sec:simu}

In this section, we verify the effectiveness of the proposed CSEPNN by experimental results.
 
All computaitons are performed using an AMD Ryzen 7 5800H with Radeon Graphics 3.20 GHz, 16GB Memory (RAM), Windows 11 (64 b) operating system, and Python $3.9$.
During the training procedure, we apply the AdamW optimizer. The initial training rate is set to $0.001$ with batch size of $1024$ and $100$ iterations. The numbers of training samples, validation samples and testing samples are $30000$, $5000$ and $4000$, respectively. $\delta$ in Huber loss is set as $\delta=0.3 \bar{y}_\star$, where $\bar{y}_\star$ denotes the mean of the labels in training samples. In addition, the grid search is applied to determine the hyperparameters. The implementation of the models is based on PyTorch as the backend and executed on the same device.

\begin{figure}[!ht]
	\centering \includegraphics[width=0.48\textwidth]{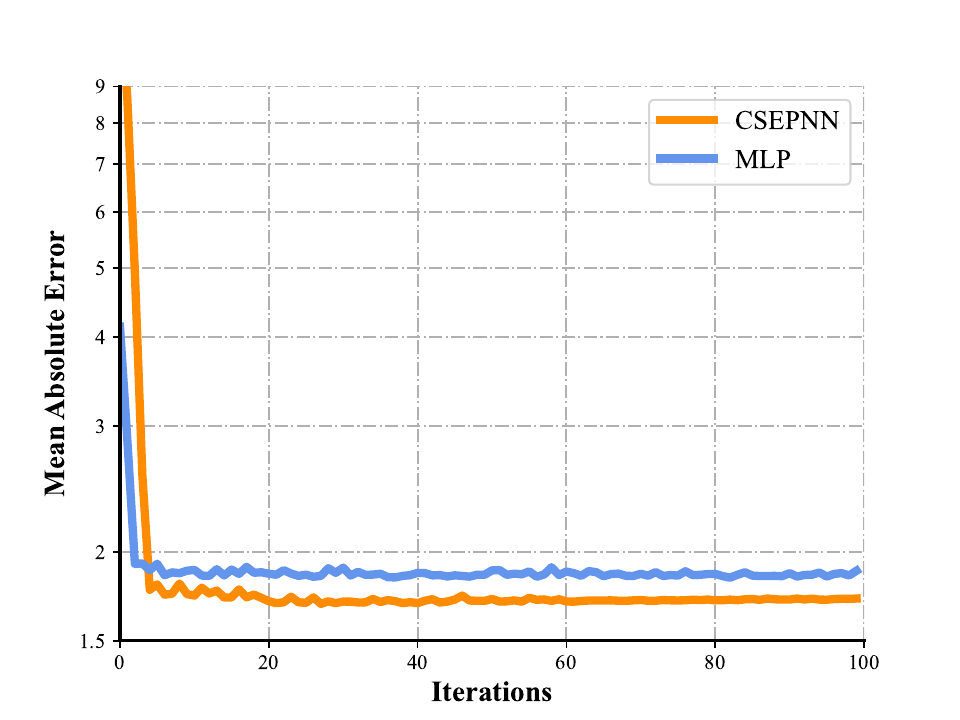}
	\centering \caption{The MAE on the validation dataset versus iterations.}
	\label{curve}
\end{figure}

Fig. \ref{curve} illustrates the MAE on the validation dataset versus the number of iterations during training. We introduce the multi-layer perceptron (MLP) as a benchmark \cite{kruse2022multi} to compare with the proposed CSEPNN. 
Note that CSEPNN presents a faster convergence performance and gives a lower bound of the achievable MAE. 
The initialized MLP model demonstrates superior initial fitting capabilities. However, after $30$ iterations, the MAE on the validation dataset of MLP remains at $2$, in contrast to the CSEPNN, which achieves a lower MAE of $1.7$. Therefore, the trained CSEPNN model exhibits a more robust ability to capture the relationship between the channel statistics and the CSE.

\begin{figure}[t]
	\centering \includegraphics[width=0.55\textwidth]{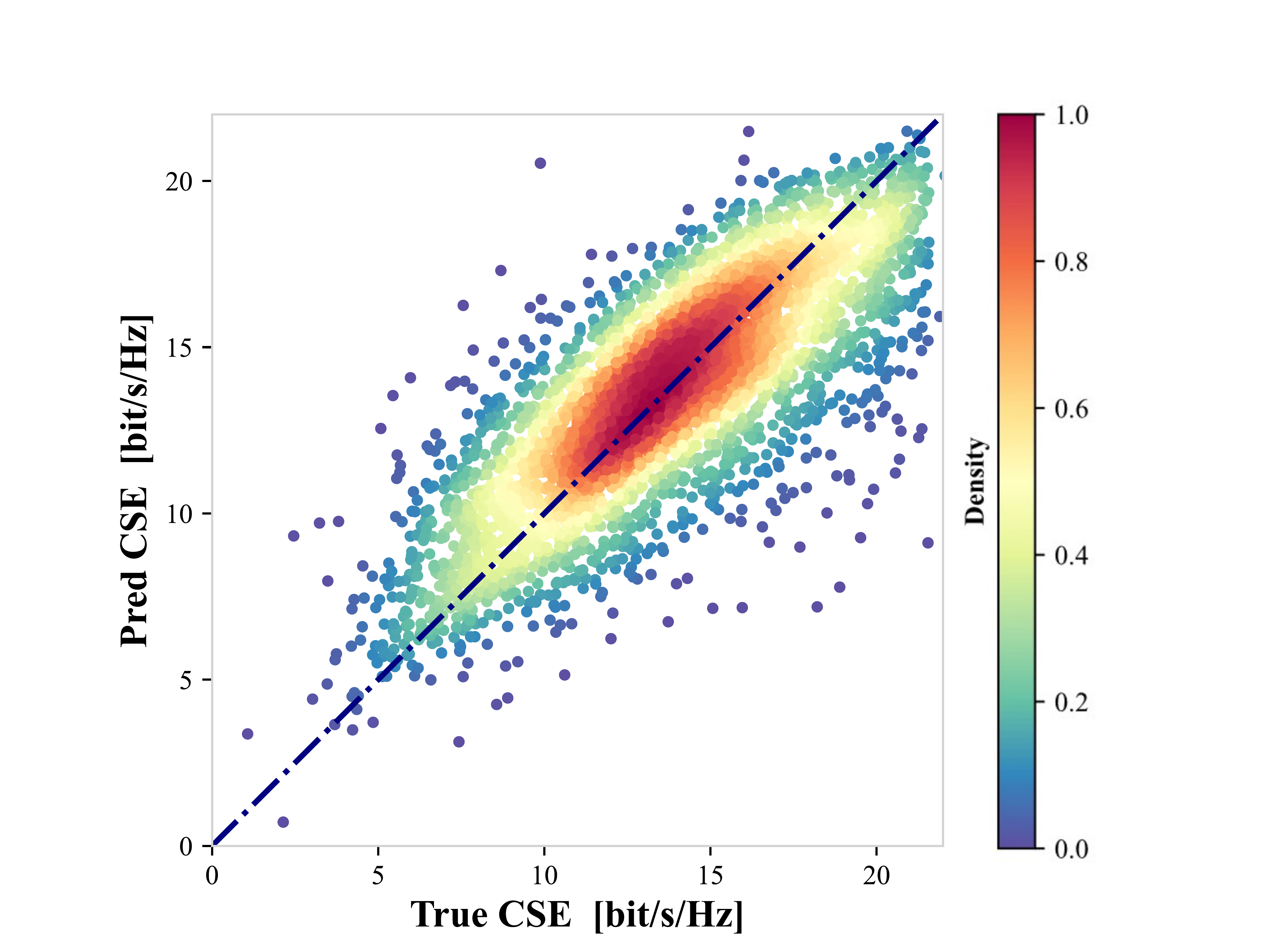}
	\centering \caption{Density scatter plot.}
	\label{DSP}
\end{figure}
\vspace{-10pt}
 Fig. \ref{DSP} depicts the predictive performance of the trained CSEPNN model on testing dataset.
A diagonal line is utilized to illustrate perfect agreement between the predicted and true CSE values.
It can be observed that a significant majority of samples cluster near the diagonal, particularly within the $12\sim15$ bit/s/Hz range, indicating high prediction accuracy. Despite the presence of some outliers with suboptimal predictions, potentially due to data cleansing oversights, these are within an acceptable error margin.

\begin{figure}[!ht]
	\centering \includegraphics[width=0.48\textwidth]{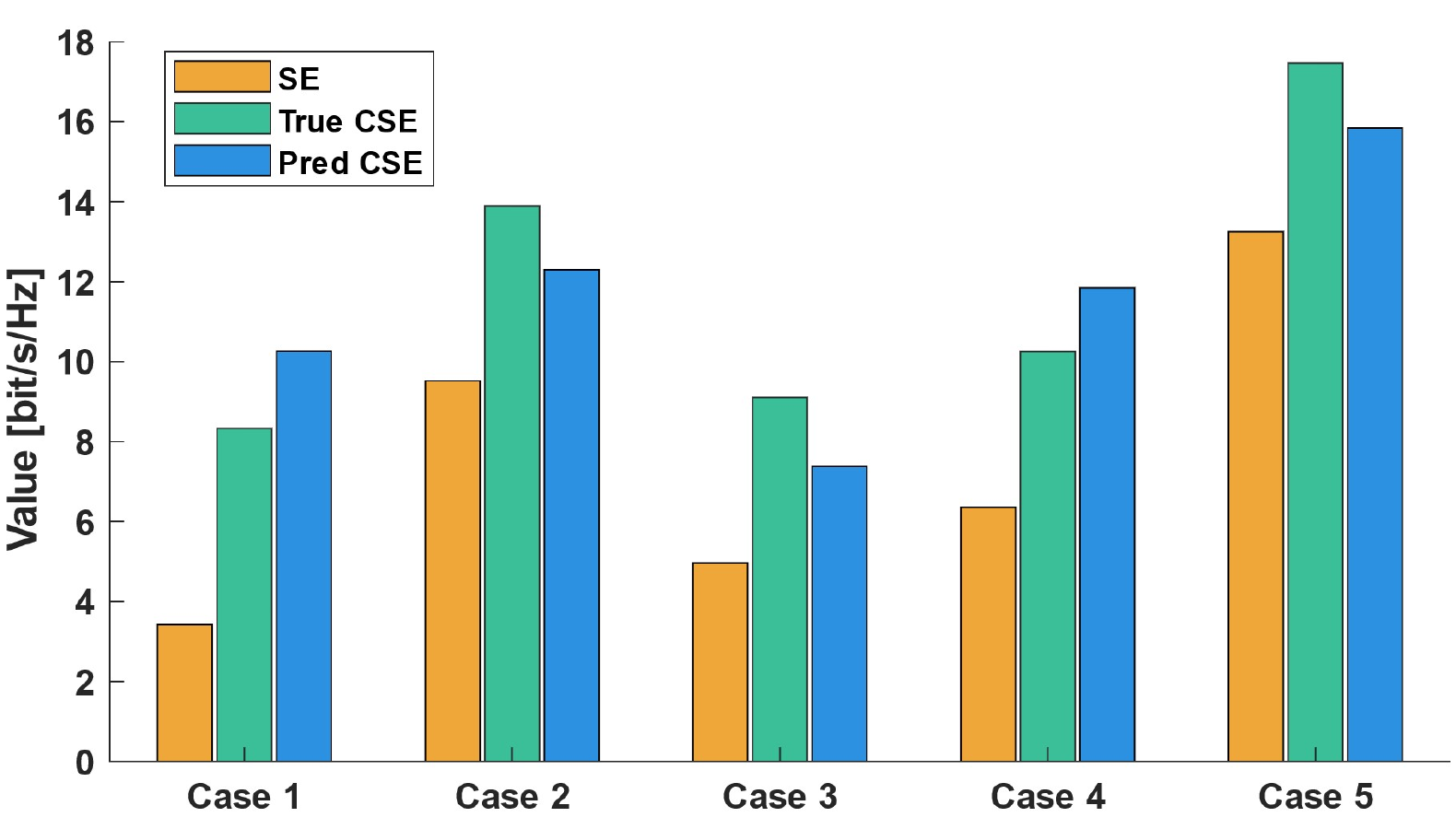}
	\centering \caption{Histogram of SE and CSE across various cases.}
	\label{histogram}
\end{figure}

 Table \ref{distribution table} presents the performance of the CSEPNN model across a range of cases.
 We evaluate five distinct datasets which are collected from different times and locations, and use MAE as performance metric. 
 As shown in  Table \ref{distribution table}, the MAE remains below $2$ for all evaluated cases.
Notably, the dataset from Case $1$  yields the highest MAE, at $1.93$. For the datasets from Case $2$, Case $4$, and Case $5$, the average MAE is approximately $1.6$.
Furthermore, Fig. \ref{histogram} shows the average SE in ST mode, the true and predicted CSE in JT mode for the aforementioned five cases.
Note that the true CSE exhibits a marked improvement over the SE, with case $1$ showing more than a twofold increase, and the predicted CSE can effectively reflect the throughput during BSs coordination.
Consequently, the CSEPNN model demonstrates high generalization across various scenarios, indicating its potential for predicting the CSE in real-world CoMP systems. 

\begin{table}[!h]
	\centering
	\caption{The performance of CSEPNN across various cases.}
	\begin{tabular}{llllll}
		\toprule
		\toprule
		& Case $1$ & Case $2$ & Case $3$ & Case $4$ & Case $5$\\
		\toprule
		\toprule
		Sample Size & 445 & 2088 & 698 &  2079& 1425\\
		True CSE (Avg) & 8.33 & 13.89 & 9.11 &10.25 & 17.47\\
		Pred CSE (Avg) & 10.26 & 12.30 & 7.38 &11.85 & 15.84 \\
		MAE & 1.93 & 1.59 & 1.73  & 1.60& 1.63 \\
		\bottomrule
	\end{tabular}
	\label{distribution table}
\end{table}

\section{Conclusion} \label{sec:conclusion}
In this letter, we have proposed a data-driven model-assisted approach to predict the CSE in real-world 5G CoMP systems. We have proposed domain knowledge-based data preprocessing and feature space definition, and then developed the CSEPNN model to capture the high-dimensional non-linear mapping function from the channel statistics to the CSE.
Experimental results demonstrate that the CSEPNN can achieve high generalization and reliable performance on real-world data. 
Future research will concentrate on real-time optimization of network parameters in CoMP systems, building upon the CSE prediction approach.

\vspace{-10pt}

\small
\bibliographystyle{ieeetr}
\bibliography{references}

\end{document}